\documentclass[fleqn,twoside]{article}
\usepackage{macros,amssymb}
\usepackage[dvips]{epsfig}
\usepackage{espcrc2}
\begin{document}
\title{
{
\vspace{-4.25cm} \normalsize \hfill
\parbox{30mm}{DESY 99-134\\OUTP 99-43P\\hep-lat/9909098}
}\\[30mm]
%
% New % low energy physics 
% applications of the QCD Schr\"odinger functional%
Low energy physics from the QCD Schr\"odinger functional%
% \thanks{Talks given by J.\ H.\ and R.\ S.\ at the International Symposium
% on Lattice Field Theory, June 29 -- July 3, 1999, in Pisa, Italy.}
\thanks{based on talks given by J.\ H.\ and R.\ S.\ at the conference
LATTICE '99, June 29 -- July 3, 1999, in Pisa, Italy}
}
\author{
Joyce Garden\address{
Department of Physics \& Astronomy, University of Edinburgh,
Edinburgh EH9~3JZ, Scotland},
Marco Guagnelli\address{
Dipartimento di Fisica, Universit\`a di Roma {\it Tor Vergata},
and INFN, Sezione di Roma II},
Jochen Heitger\address{
Deutsches Elektronen-Synchrotron DESY Zeuthen, Platanenallee~6,
D-15738 Zeuthen, Germany},
Rainer Sommer$^{\mbox{\scriptsize c}}$
and Hartmut Wittig\address{
Theoretical Physics, University of Oxford, 1~Keble Road, 
Oxford OX1~3NP, UK}%
\thanks{PPARC Advanced Fellow}\\
\vskip 0.25cm
(ALPHA and UKQCD Collaborations)
}

\begin{abstract}
We review recent work by the ALPHA and UKQCD Collaborations 
where masses and matrix elements were computed in lattice
QCD using Schr\"odinger functional boundary conditions and 
where the strange quark mass was determined in the quenched approximation.
We emphasize the general concepts and 
our strategy for the computation of quark masses. 
\end{abstract}

\maketitle

\section{Introduction}
Based on the recent progress in lattice QCD concerning non-perturbative
renormalization of local composite operators and removal of leading
discretization errors \cite{Kuramashi:lat99,tsuk97:rainer,lat97:hartmut},
there is a natural interest to calculate hadron masses and matrix elements
in the continuum limit.
Especially, the problem of a determination of the light quark masses
\cite{Kenway:lat98} can be addressed with confidence, since the complete
renormalization is known non-perturbatively~\cite{mbar:pert,mbar:pap1}.

As an alternative to more standard methods, we will show that the
Schr\"odinger functional (SF) framework allows for a
reliable computation of spectral quantities with high accuracy in
lattice QCD.
As a necessary prerequisite, we will demonstrate explicitly that
SF correlation functions are dominated by hadron
intermediate states at Euclidean time separations
of around $3\,\fm$.

This technique is then applied to physical observables in the meson
sector of QCD, in particular to light quark masses.
We present an overview of the work discussed in
refs.~\cite{mbar:pap2,mbar:pap3}.
All results refer to the fully $\Oa$ improved theory, but
due to the rather introductory character all issues 
related to $\Oa$ improvement are ignored here.

Our quenched data stem
from four lattices (SF boundary conditions) of 
spatial extent $L\approx1.5\,\fm$ and time extension
$T = 2L$ with lattice spacings, $a$, from 0.1 fm ($\beta=6.0$) to 0.05 fm
($\beta=6.45$), and with four values of the quark mass each, so that all
results can be extrapolated first in the pseudoscalar mass and finally to
the continuum limit. For our choice of parameters,
chiral perturbation theory predicts negligible
finite size effects and an explicit investigation \cite{mbar:pap3},
which is not discussed here, confirms this.

\section{SF correlators at large time separations}
The SF is defined as the QCD partition function in a
$L^3\times T$ cylinder with periodic boundary conditions in three of the
four Euclidean dimensions and Dirichlet boundary conditions in time at
the hypersurfaces $x_0=0$ and $x_0=T$ \cite{SF:LNWW}.
Whereas in earlier applications, e.g.~in refs.~\cite{mbar:pert,mbar:pap1},
correlation functions in the SF were mainly considered in the perturbative
regime (i.e.~small extensions of the space-time volume), the emphasis in
this section is on their properties in intermediate to large volumes with
extensions significantly larger than the typical QCD scales of order
$1~\fm$.

Starting from the quantum mechanical interpretation of the SF
\cite{SF:LNWW,SF:stefan1}, one can derive explicit expressions for the
representation of its correlation functions in terms of intermediate
physical states \cite{mbar:pap2}.
Let us discuss an example which is directly relevant in section 3.
We consider the correlation functions 
\bes
  \fx(x_0)  & = & - \frac{L^3}{2}\langle X(x) \, \op{} \rangle \,,
  \label{e_fa} \\
  f_1       & = & - \frac{1}{2}\langle \op{}' \, \op{} \rangle 
  \label{e_f1} \,,
\ees
where $\op{}$ is a {\ps} field, constructed from a $\vecp=0$
quark field and a $\vecp=0$ antiquark field at $x_0=0$ :
$\op{}={{a^{6}}\over{L^3}}\sum_{\vecy,\vecz}
\zetabar_{i}(\vecy)\gamma_5\zeta_{j}(\vecz)$\,,
$i,j$ being flavor indices.  
Analogously, $\op{}'$ is located at the boundary $x_0=T$. 
One may think of the fields  $\zeta,\ldots\,$ as quark fields 
at the boundary; their precise definition is given in \cite{impr:pap1}.
In the following,
we shall be interested in $\fa$, defined by
$X(x)=A_0(x)=\psibar_{j}(x)\gamma_0\gamma_5\psi_{i}(x)$, 
and $\fp$, given by $X(x)=P(x)=\psibar_{j}(x)\gamma_5\psi_{i}(x)$.

For large separations $x_0$ and $T-x_0$ the correlation functions
are dominated by the lowest lying intermediate states with the appropriate 
quantum numbers. These are the pseudoscalar ground state and the 
vacuum, the latter contributing between $x_0$ and $T$ in both 
$\fa$ and $\fp$.
In the transfer matrix formalism one may then arrive \cite{mbar:pap2} 
at the asymptotic behaviour including the first non-leading corrections:
\bes
  \fx(x_0) &\approx& \frac{L^3}{2}\,\rho\,
                     \langle 0,0|\opX|0,{\rm PS}\rangle   
                     \,\rme^{-x_0 \mp } \nonumber \\
           &       & \times\left\{ 1 + \etax^{\rm PS}\rme^{-x_0 \Delta } 
                     +\etax^{0} \rme^{-(T-x_0) m_{\rm G} } \right\} 
  \label{e_fa_asympt} \nonumber \\
  f_{1}    &\approx& \frac{1}{2}\,\rho^2 \, \rme^{-T \mp } \,.
  \label{e_f1_asympt}
\ees
In leading order an unknown matrix element $\rho$ and the desired
matrix elements of the Schr\"odinger picture operators 
$\opX$, associated with the fields $X$, arise. 
The mass of the $0^{++}$ glueball, $m_{\rm G}$, and the gap in the
pseudoscalar channel, $\Delta$, enter the corrections with
coefficients $\etax^{0},\etax^{\rm PS}$ expressed through certain matrix
elements.
Apart from the pseudoscalar mass, $\mp$, the formulae (\ref{e_f1_asympt})
enable to calculate the decay constant, $\Fp$, and the
pseudoscalar coupling, $\Gp$, via
\bes
  \langle 0,0|\opA|0,{\rm PS}\rangle & = & \Fp\mp(2\mp L^3)^{-1/2} \,,\\
  \langle 0,0|\opP|0,{\rm PS}\rangle & = & \Gp(2\mp L^3)^{-1/2}    \,.
\ees
Note in particular that $f_{1}$ may be used to eliminate the
unknown factor $\rho$, the only place 
where the divergent renormalizations
of the boundary quark fields are hidden. 
A generalization e.g.~to the vector meson channel is straightforward.

\begin{figure}[t]
\begin{center}
\epsfig{file=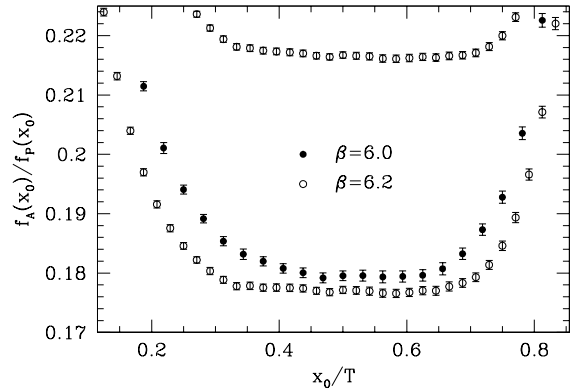,width=7.5cm}
\vspace{-1.2cm}
\caption[t_param]{\label{PlatPlot}
                  The ratio ${\fa/\fp}$ for 
                  $a\approx0.07\,\fm$ (open circles) and 
                  $a\approx0.09\,\fm$ (filled circles).}
\end{center}
\vspace{-0.8cm}
\end{figure}

\begin{figure}[t]
\begin{center}
\epsfig{file=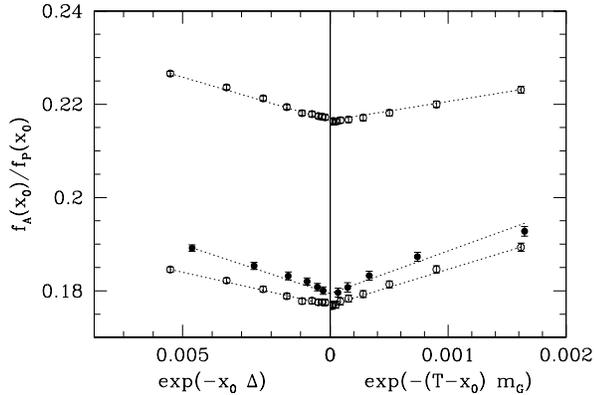,width=7.5cm}
\vspace{-1.2cm}
\caption[t_param]{\label{ExstPlot}
                  Influence of excited state contributions on 
                  ${\fa/\fp}$ at the same parameters as in
                  \fig{PlatPlot}.}
\end{center}
\vspace{-0.8cm}
\end{figure}

We turn to the numerical tests of the practicability of our method.
As a representative example, which moreover will play a crucial r\^{o}le
in the next section, we discuss the ratio
\[
  {\fa \over \fp} =
  {\mp\Fp \over \Gp }\left\{
  1+\rmO\left(\rme^{-x_0 \Delta },\rme^{-(T-x_0)m_{\rm G}}
  \right) \right\}\,.
\]
It shows approximate plateaux between $x_0 \approx 1\,\fm$ and
$T-x_0 \approx 1\,\fm$ (\fig{PlatPlot}).
%
%%% Beginn Figur %%%
%%% Ende Figur %%%
%
Their height determines $\mp\Fp/\Gp$. 
Before, however, just reading these off, the magnitude of the
``contaminations'' by excited state contributions should be assessed. 
Given the fact that estimates for $\Delta$ and $m_{\rm G}$ are 
available \cite{mbar:pap2}, this may be achieved
by plotting the ratio against the leading corrections:
the slopes in \fig{ExstPlot} allow to deduce
the range $t_{\rm min}\leq x_0 \leq t_{\rm max}$
where the excited state corrections are below a certain systematic
error margin, which we chose in turn to be 
negligible compared to the final statistical errors. 
The ratio was then averaged in the range
$t_{\rm min}\leq x_0 \leq t_{\rm max}$.
%
%%% Beginn Figur %%%
%%% Ende Figur %%%
%
%
%%% Beginn Figur %%%
\begin{figure}[t]
\begin{center}
\epsfig{file=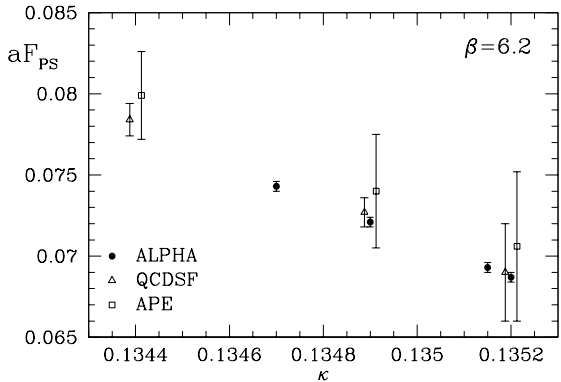,width=7.5cm}
\vspace{-1.2cm}
\caption[t_param]{\label{CompPlot}
                  Comparison of some of our results (ALPHA) with those
                  of the QCDSF \cite{impr:qcdsf} and APE
                  \cite{impr:roma1_spec} Collaborations.
                  Symbols are slightly displaced.}
\end{center}
\vspace{-0.8cm}
\end{figure}
%%% Ende Figur %%%
%
This procedure is easily extended to the calculation of $\mp,\Fp$ and
$\Gp$ separately.

The numerical efficiency of the computations with SF correlation
functions is illustrated in \fig{CompPlot}, where we compare 
our results for the bare (improved)
pseudoscalar decay constant with those found in the literature.
To judge the relative size of the statistical errors 
one should note that our statistics is 
not much higher than the statistics present in the 
other simulations.

Clearly, the SF applied for extracting hadron physics is 
similar to the method of wall sources, but one should note
that gauge invariance is kept
at all stages of the present formulation. Furthermore,
dimensionless non-local fields are used to create the boundary
states. Dimensional analysis then tells us that the pre-asymptotic
decay of SF correlation functions is slow, leaving large and precise
signals at hadronic length scales of 1-2 fm.
Since the correlation functions are renormalizable by simple factors,
this property is indeed independent of the lattice spacing, once one
is sufficiently close to the continuum limit.

\section{Quark masses: the strategy}
Before we come to the numerical applications of the above methods, we 
want to explain in detail our strategy for the computation of
light quark masses.
Their ratios can be computed from chiral perturbation
theory~\cite{reviews:quarkmasses}.
The currently best results read
\bes
  M_\up/M_\down = 0.55\pm0.04\,,\,\, 
  M_\strange/\Mlight = 24.4 \pm 1.5  
  \label{e_ratios}
\ees
with $\Mlight=\frac12(M_\up+M_\down)$ \cite{leutwyler:1996}.

Despite this success, there is substantial work to be done using lattice QCD.
\begin{itemize}
\item
   The applicability of chiral perturbation theory needs to be checked. This 
   concerns the practical question in how far the lowest orders dominate the 
   full result. 
\item
   The parameters in the  chiral Lagrangian (at a given order in the
   expansion) can not be inferred with great precision from
   experimental data alone.
   Their determination can be improved significantly by help of lattice QCD 
   results.    
\item
   In particular, there is one parameter in the chiral Lagrangian which is 
   impossible to determine from experimental data. This is the overall scale 
   of the quark masses, which is only defined once the connection with 
   the fundamental theory, QCD, is made. 
\end{itemize}
An important point to realize is that all of the above problems can be dealt
with by working with {\em unphysical quark masses} (of course they should not 
be too large). For the first two problems it is in fact essential to 
explore a range of quark masses, and also for the last problem, 
which we address here, this is of significant advantage. 

The above considerations lead us to determine an implicitly defined
reference quark mass $\Mref$,
\bes 
  \mp^2(\Mref)\rnod^2 = \rrmm        =(\mk \rnod)^2 \,,
\label{mref_def}
\ees
where $\mp^2(M)$ is the pseudoscalar meson mass as a function
of the quark mass (mass-degenerate quarks).
Chiral perturbation theory {\em in full QCD} relates $\Mref$ to the
other light quark masses viz. 
\bes 
  2\Mref \approx M_{\strange} + \Mlight \,.
\label{mref_ms}
\ees
Further evidence for this relation is given in \cite{mbar:pap3}, using the 
results of quenched lattice QCD.

What was said above
is valid literally in mass-independent renormalization schemes, where quark
masses are renormalized with a flavor independent factor. 
The resulting running 
quark masses, $\mbar(\mu)$, are however 
scheme- and scale-dependent quantities. It is
advantageous to compute directly the 
renormalization group invariant (RGI) quark masses \cite{mbar:pap1},
which are pure numbers and do
not depend on the scheme. In terms of $\mbar(\mu)$ they are 
given as
\bes
 M &\equiv& \lim_{\mu\to\infty} \left\{(2b_0\gbar^2(\mu))^{-d_0/2b_0}\, 
 \mbar(\mu)\right\} \,, \label{e_M_i1}\\
 &&b_0=11/(4\pi)^2,\; d_0=8/(4\pi)^2 \,.
\ees
For $\Oa$ improved quenched QCD,
the ALPHA Collaboration has determined the renormalization 
factor $\zM$ relating
the bare current quark masses $m$, defined by the PCAC relation,  to the 
RGI masses \cite{mbar:pert,mbar:pap1}:
\bes
M = \zM m \,.
\ees
Applying the PCAC relation to the vacuum-to-pseudoscalar matrix elements
then results directly in our central relation,
\bes
  {2 r_0 \Mref} & = & \zM {\Rwi|_{r_0^2\mp^2 =\rrmm} \over \,r_0} 
                    \times \rrmm \,,
  \label{e_M_r0_def} \\
  R           & = & {\Fp / \Gp}\,.
  \label{e_defR}
\ees

\section{Results}
\begin{figure}[t]
\begin{center}
\vspace{-1.8cm}
\epsfig{file=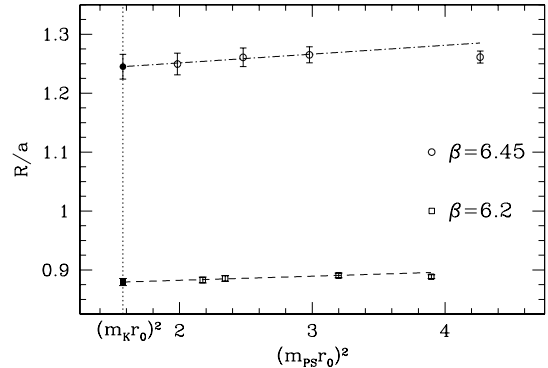,width=7.5cm}
\vspace{-1.8cm}
\caption[t_param]{\label{ChirPlot}
                  Mass dependence and extrapolations at
                  the two smallest values of the lattice
                  spacing.}
\end{center}
\vspace{-0.8cm}
\end{figure}

\begin{figure}[t]
\begin{center}
\epsfig{file=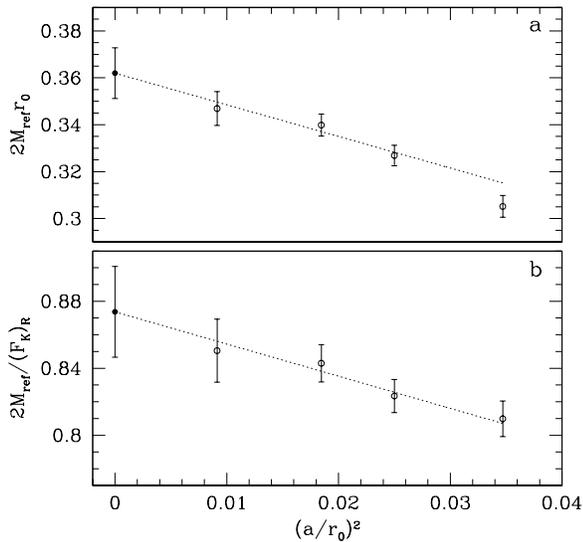,width=7.5cm}
\vspace{-1.5cm}
\caption[t_param]{\label{ContPlot}
                  Continuum extrapolations of $\Mref$ in units of
                  $r_0$ (a) and $(\Fk)_{\rm R}$ (b). 
                  Full symbols are the $a=0$ limits.
                  The (dashed) fit functions are continued
                  beyond their range towards larger $a$.}
\end{center}
\vspace{-0.8cm}
\end{figure}

With the methods of section~2, the ratio $R/a$ and the 
meson mass $\mp a$ can be computed accurately as a 
function of the bare quark mass and the bare coupling.
Also the scale $\rnod/a$ is known \cite{pot:r0_SU3}.  
A mild extrapolation, shown in \fig{ChirPlot}, then yields
$R/a$ at the point $r_0^2\mp^2 =\rrmm$.\footnote{
   The reason for not computing directly at or 
   below the quark mass corresponding to $r_0^2\mp^2 =\rrmm$ 
   was that this is the region where unphysical zero modes 
   of the $\Oa$ improved Dirac operator are relevant 
   (``exceptional configurations''). It is now
   understood that this technical problem can be resolved 
   by introducing the quark mass into the lattice theory
   in a particular way \cite{TWQCD:lat99}. However, even
   without entering this region, our quark mass
   points are close enough to perform a safe extrapolation.
} 
Next the full combination $r_0 \Mref$ 
(see eqs.~(\ref{mref_def},\ref{mref_ms})) is extrapolated to
the continuum limit in \fig{ContPlot}a. 
Since this latter extrapolation involves a significant slope
we discarded the point furthest away from the continuum. This 
is a safeguard against higher order lattice spacing effects. 

Furthermore, the entire analysis was repeated for 
$\Mref$ in units of the Kaon decay constant \cite{mbar:pap3}.
In this case the lattice spacing dependence is
weaker (\fig{ContPlot}b).
The two results, 
\bes  
  2\, r_0 \Mref = 0.36(1)  \nonumber \\
   \stackrel{\rnod=0.5\,\fm}{\longrightarrow} 
   &2\, \Mref = 143(5)\,\MeV \,,& 
  \nonumber \\
  {2\, \Mref / (\Fk)_{\rm R}} = 0.87(3) \nonumber \\
    \stackrel{(\Fk)_{\rm R}=160\,\MeV}{\longrightarrow}  
  &2\, \Mref = 140(5)\,\MeV\,,&  \nonumber 
\ees
agree better than one would expect them to do in the quenched 
approximation.
%
%%% Beginn Figur %%%
%%% Ende Figur %%%
%
%
%%% Beginn Figur %%%
%%% Ende Figur %%%
%

It is premature to conclude that this approximation yields a nearly
unique result.
Rather, the assignment of physical units {\em is} ambiguous.
One may estimate \cite{mbar:pap3} by help
of the recent results of the CP-PACS Collaboration \cite{qspect:CPPACS}
that roughly $10\%$ larger numbers would be obtained, if the 
scale $\rnod$ were replaced by one of the masses 
of the stable light hadrons.  This represents a typical ambiguity 
of the quenched approximation.

Other interesting results after extrapolation to the continuum limit are
\[
  \left\{\rnod (\Fp)_{\rm R}\right\}_{r_0^2\mp^2 =\rrmm} =
  0.415(9)
\]
compared to $0.5\,\fm \times (\Fk)_{\rm R} = 0.405(5)$ with the
experimental value of $(\Fk)_{\rm R}$, and
\[
  \left\{\rnod m_{\rm vector} \right\}_{r_0^2\mp^2 =\rrmm} =
  2.39(7)
\]
to be confronted with $0.5\,\fm \times \mkstar=2.26$. In the latter
comparison one should remember that the K$^*$ meson is unstable
even in absence of electroweak interactions.
Its width, $\Gammakstar$, amounts to $0.5\,\fm \times \Gammakstar=0.13$ .

\section{Discussion}
In addition to the -- by now well known  -- applications
to renormalization problems in QCD, the \SF has been shown to be 
useful for the computation of low energy matrix elements.
For the particular cases of vacuum-to-{\ps} matrix elements, 
our results are significantly more precise than those from 
standard methods.
This implies a precise computation (in the continuum limit) 
of the renormalization group
invariant quark mass $\Mref$ defined above. 
$\MS$ masses for finite renormalization scales $\mu$ are obtained
through perturbative conversion factors known up to 4-loop precision.
A typical result is
\[
  \mbar_{\strange}(2\,\GeV) =97(4)\,\MeV\,,
\]
where the uncertainty in $\Lambda_{\msbar}^{(0)}=238(19)\,\MeV$
\cite{mbar:pap1}, entering the relation of the running quark masses in
the $\msbar$ scheme to the RGI
masses, has been accounted for and the quark mass ratios
from full QCD chiral perturbation theory were used.

One should keep in mind that there is an intrinsic ambiguity when assigning
physical units in the quenched approximation. 
We estimated this to be of order $10\%$.

%% This work is part of the ALPHA collaboration research programme.
%% We thank DESY for allocating computer time to this project.

%

%

\begin{thebibliography}{99}

\bibitem{Kuramashi:lat99}
Y. Kuramashi,
\newblock these proceedings.

\bibitem{tsuk97:rainer}
R. Sommer,
\newblock Nucl. Phys. Proc. Suppl. 60A (1998) 279, hep-lat/9705026.

\bibitem{lat97:hartmut}
H. Wittig,
\newblock Nucl. Phys. Proc. Suppl. B63 (1998) 47, hep-lat/9710013.

\bibitem{Kenway:lat98}
R.D. Kenway,
\newblock Nucl. Phys. Proc. Suppl. B73 (1999) 16, hep-lat/9810054. 

\bibitem{mbar:pert}
S. Sint and P. Weisz,
\newblock Nucl. Phys. B545 (1999) 529, hep-lat/9808013.

\bibitem{mbar:pap1}
S. Capitani, M. {L\"uscher}, R. Sommer and H. Wittig,
\newblock Nucl. Phys. B544 (1999) 669, hep-lat/9810063.

\bibitem{mbar:pap2}
M. Guagnelli, J. Heitger, R. Sommer and H. Wittig,
\newblock (1999), hep-lat/9903040.

\bibitem{mbar:pap3}
J. Garden, J. Heitger, R. Sommer and H. Wittig,
\newblock (1999), hep-lat/9906013.

\bibitem{SF:LNWW}
M. {L\"uscher} et~al.,
\newblock Nucl. Phys. B384 (1992) 168, hep-lat/9207009.

\bibitem{SF:stefan1}
S. Sint,
\newblock Nucl. Phys. B421 (1994) 135, hep-lat/9312079.

\bibitem{impr:pap1}
M. {L\"uscher}, S. Sint, R. Sommer and P. Weisz,
\newblock Nucl. Phys. B478 (1996) 365, hep-lat/9605038.

\bibitem{impr:qcdsf}
M. {G\"ockeler} et~al.,
\newblock Phys. Rev. D57 (1997) 5562, hep-lat/9707021.

\bibitem{impr:roma1_spec}
D. Becirevic et~al.,
\newblock (1998), hep-lat/9809129.

\bibitem{reviews:quarkmasses}
J. Gasser and H. Leutwyler,
\newblock Phys. Rept. 87 (1982) 77.

\bibitem{leutwyler:1996}
H. Leutwyler,
\newblock Phys. Lett. B378 (1996) 313, hep-ph/9602366.

\bibitem{pot:r0_SU3}
M. Guagnelli, R. Sommer and H. Wittig,
\newblock Nucl. Phys. B535 (1998) 389, hep-lat/9806005.

\bibitem{TWQCD:lat99}
R. Frezzotti, P.A. Grassi, S. Sint and P. Weisz,
\newblock these proceedings, hep-lat/9909003.

\bibitem{qspect:CPPACS}
S. Aoki et~al.,
\newblock (1999), hep-lat/9904012.

\end{thebibliography}
\end{document}